# Insights into the Evolution of Horizons from Non-Orthogonal Temporal Coordinates[1]

James Lindesay[2]
Computational Physics Laboratory
Howard University,
Washington, D.C. 20059

Abstract

The introduction of coordinates representing the points of view of various observers results in the possibility of horizons when acceleration and gravitation are included.  A horizon is a surface of possible light beams in a region of space of finite distance from the observer, which means that since nothing travels faster than light, events on the far side of a horizon cannot influence those on the causal side.  A black hole has such a horizon, where some radially outgoing light beams can never reach a distant (or even nearby) observer.  However, since one suspects that black holes can swallow energy, and even evaporate by Hawking radiation, such horizons must take on a time dependency.  A naive introduction of temporal dependency results in infinities (singularities) in energy densities, suggesting in such descriptions that an in-falling observer would encounter a hard surface at the horizon.  However, if coordinates representing space-time as analogous to a "flowing river" are used to describe the dynamics of a black hole, no such singularities are encountered.  Such a parameterization of time dependent horizons will be offered in this presentation.  A Penrose space-time diagram (which represents the entire space-time on a finite diagram with light beams always moving at a 45 degree angle to vertical) describing the growth and evaporation of an example black hole, along with the resulting coordinate anomaly, will be constructed.

I. Introduction

One of the principles of modern physics that is most adhered to is that one expects that the models that we construct to describe the phenomena of the physical universe should not depend upon any absolute frame of reference.  The

---

discovery of the Cosmic Microwave Background radiation perhaps demonstrates a counter example to this supposition, due to the preferred frame at rest relative to the energy content of the universe during its "initial" phase of expansion. However, for most phenomena, the co-variance of the laws modeling those phenomena is consistent with the expectation of an independence of the particular frame of reference utilized by the observer. This principle is embodied in the concept of *complementarity*[1] in the description of black holes. In its most direct expression, complementarity simply states that no observer should ever witness a violation of a law of nature. In particular, one expects that for a freely falling observer, there should be no local affects of gravitation as espoused by the principles of equivalence and relativity.

A naive introduction of a dependency of the mass of a black hole on the Schwarzschild time coordinate results in singular behavior of curvature invariants at the horizon, violating expectations from complementarity. A singularity in a curvature invariant defines singular behavior in the physical energy content through Einstein's equation, implying that an in-falling observer might (figuratively) encounter a "brick wall" at the horizon. If instead a temporal dependence is introduced in terms of a coordinate akin to the river time representation[2], the Ricci scalar is nowhere singular away from the origin. It is found that for a changing mass scale due to accretion or evaporation, the null radial geodesics that generate the horizon are slightly displaced from the coordinate anomaly. In addition, a changing horizon scale significantly alters the form of the coordinate anomaly in diagonal (orthogonal) metric coordinates representing the space-time. We examine black hole evolution using coordinates that introduce no singularities away from the origin, and construct a Penrose diagram describing the growth and evaporation of an example black hole, along with the resulting coordinate anomaly.

II. Special Relativity and Space-Time Diagrams

The consistency of the laws of electromagnetism as described in Maxwell's equations resulted in the development of the special theory of relativity. Maxwell's equations predict a frame-independent speed of electromagnetic radiation including light. Since speed measures relative distance per time, and since distance is relative, then time must likewise be relative in order to guarantee the constancy of the speed of light. One then must develop a standard for the construction of a space-time grid to describe global coordinates. In special relativity, this is usually done using the following steps[3]:

- Set up a pre-arranged set of stop-clocks at appropriate locations and preset clock times corresponding to the expected light travel times.
- Construct each stop-clock to initiate standard tick rates after receiving a photon from an initiating light pulse set up to synchronize the clocks.

- Initiate each stop-clock when the light pulse beam reaches each individual clock.
- The clocks are all then synchronized once they are running. Once the light pulse has passed all of the standard clocks, they are all running, and they all read the same time.

The construction of space-time coordinate grids allows us to set up experiments with various perspectives corresponding to various observers. The various events can be represented on space-time diagrams. An example of such a space time diagram is given in Figure 1.

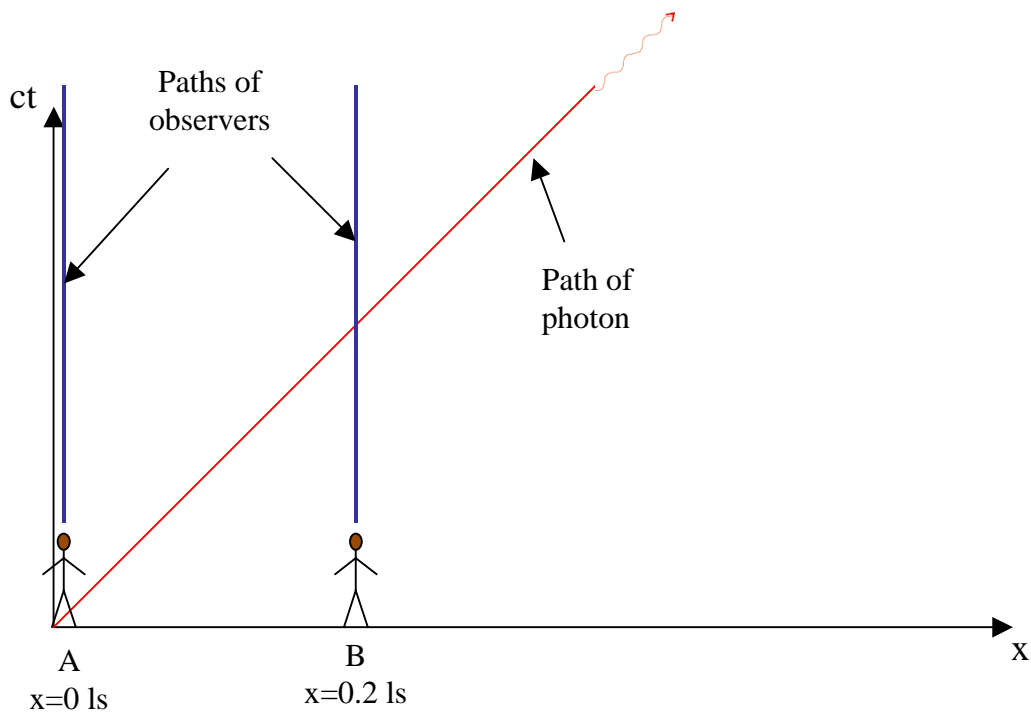

Space-time diagram

In this figure, observer A at the origin is stationary, indicated by no spatial (x) displacement and only temporal (ct) displacement. The observer B is likewise stationary, located a distance 0.2 light-seconds to the right of observer A. A photon sent by observer A at time t=0 that moves to the right reaches observer B at time $t_B$=0.2 seconds.

This global space-time parameterization that satisfies the invariance of the speed of light regardless of the motion of the observer is known as Minkowski space-time. The invariance of the speed of light is maintained by requiring that the space-time distance traveled by a light pulse must vanish regardless of the inertial coordinate grid utilized for observation. The invariant proper length and proper time in this *flat* space-time is given by

$$ds^2 = -c^2 dt^2 + dx^2 + dy^2 + dz^2 = -c^2 d\tau^2.$$

)The squared temporal displacement in special relativity is seen to have a sign metric opposite that of the squared spatial displacement. The form of this metric insures that photons follow null space-time trajectories (geodesics) with a constant speed c, regardless of the motion of the observer:

$$ds^2 = 0 \quad \Rightarrow \quad \frac{dx_\gamma}{dt} = c.$$

## Minkowski space-time Penrose diagram

The space-time diagram demonstrated in Figure 1cannot represent the large scale structure of Minkowski space-time, since large distances and times cannot be fit on the finite page. Penrose diagrams are convenient for diagrammatically studying the large scale structure of space-time. Penrose diagrams map infinite space-time coordinates onto a finite page, while preserving the flat space-time slope of light-like curves. Because light-like curves have slopes that are 45 degrees from the vertical, one can directly examine potential causal relationships between events occurring in disparate regions of the global space-time. Generally, one can choose any function (like hyperbolic tangents) that maps infinite arguments into finite values to insure that all points in space-time can be contained in a finite diagram. One uses *conformal* coordinates as arguments of the function in order to insure that light beams are generally represented by lines with 45 degree slopes relative to the vertical. Light beams parameterized in terms of conformal coordinates traverse equal spatial coordinate displacements in equal temporal coordinate intervals. Since this is already true for Minkowski space-time, no further transformation is necessary. The coordinate labeling the horizontal axis representing varying spatial displacements from the center $r=0$ is given by

$$Y_\rightarrow = \frac{-Tanh(ct-r)+Tanh(ct+r)}{\sqrt{2}},$$

while the coordinate labeling the vertical axis representing varying positive and negative temporal displacements is given by

$$Y_\uparrow = \frac{Tanh(ct-r)+Tanh(ct+r)}{\sqrt{2}}.$$

This coordinate transformation is shown in Figure 2.

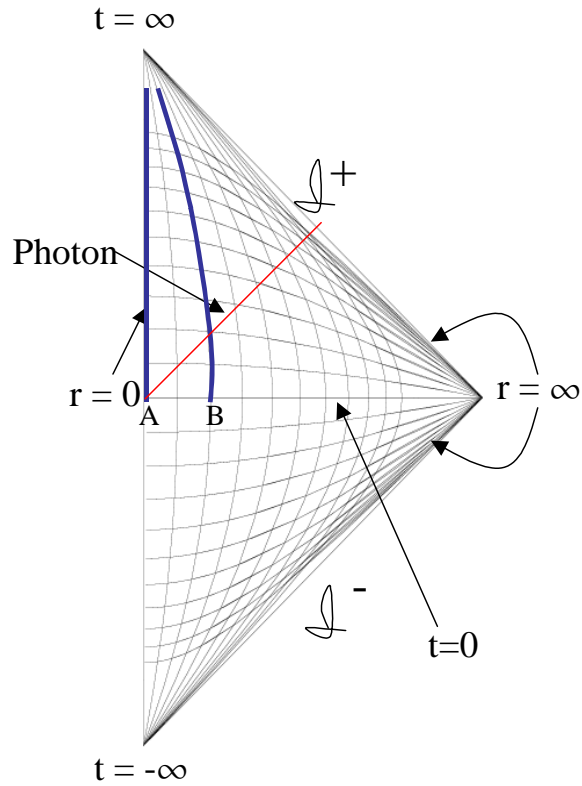

Minkowski space-time Penrose diagram As in Figure 1, the coordinates are expressed in units of tenths of light seconds The world-lines of the observers A and B, as well as that of the photon, are labeled in the Penrose diagram. The outgoing photon is seen to cross equal r and ct coordinates. There are no causally disconnected regions in Minkowski space-time, since any photon can eventually cross any spatial coordinate.

III.   Horizons

There are some space-time geometries that have causally disconnected regions, in the sense that some light beams originating in one region will never reach another region. As an example of a coordinate set that demonstrates a horizon, consider an accelerating rocket in Minkowski space-time that maintains a constant proper acceleration a. This means that any observer at rest inside the rocket must maintain a force m a as an artificial weight in order to maintain his or her constant acceleration. This artificial gravity embodies the principle of equivalence. The coordinates of the rocket as measured by a stationary inertial frame of reference can be obtained by examining the successive Lorentz transformations into the momentarily co-moving inertial frames that track the rocket's motion. The time $t$, spatial position $x$, speed of the rocket $v$ and Lorentz factor $\gamma$ as measured by the stationary inertial observer can be parameterized in terms of the proper time $\tau$ as measured by a clock in the rocket. These parameters are given by

$$t = \frac{c}{a}\sinh\left(\frac{a\tau}{c}\right) \quad , \quad \frac{v}{c} = \tanh\left(\frac{a\tau}{c}\right)$$

$$\gamma = \cosh\left(\frac{a\tau}{c}\right) = \sqrt{1 + \left(\frac{a\tau}{c}\right)^2}$$

$$x = \frac{c^2}{a}\left[\cosh\left(\frac{a\tau}{c}\right) - 1\right] = \frac{c^2}{a}\left[\sqrt{1 + \left(\frac{a\tau}{c}\right)^2} - 1\right]$$

As an example, Figure 3 illustrates the space-time trajectory of a rocket that accelerates away from the Earth with an acceleration of 1 g, such that the occupants of the rocket will feel their normal weights during the trip. The slope of the rocket trajectory is seen to approach 45º associated with the speed of light within a couple of years.

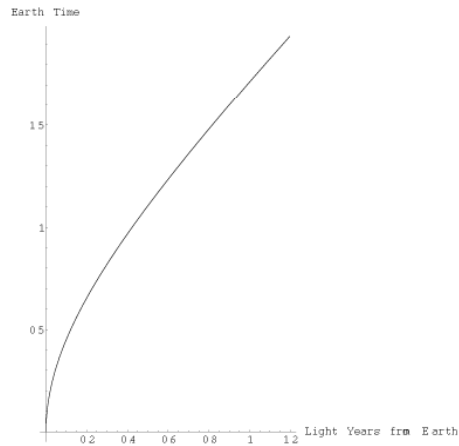

Rocket accelerating at 1 g

However, since the rocket's speed asymptotically approaches that of light, this means that there are some photons to the left of the figure that will *never* reach the rocket. We can invert Equation (4) to determine the proper time that it takes for a photon emitted from the front and rear of the rocket to reach the middle of the rocket a proper distance L away. The proper times measured for the photon to travel this proper distance are given by

$$\tau_{photon}^{from\ front} = \frac{c}{a}\log\left(1+\frac{aL}{c^2}\right)$$

(5a)

$$\tau_{photon}^{from\ rear} = -\frac{c}{a}\log\left(1-\frac{aL}{c^2}\right)$$

From Eq. (5b), one sees that a photon traveling from the rear of the rocket with an initial proper distance $L \geq c^2/a$ will **never** reach the front! This corresponds to a horizon for the coordinates established by observers in the rocket.

Therefore, there is a region of the universe that can never be accessible to observation by those in the accelerating spaceship.  The path of the first set of photons that can never reach the spaceship defines the *horizon* of observers in the spaceship.  This is demonstrated by the dashed line in Figure 4.

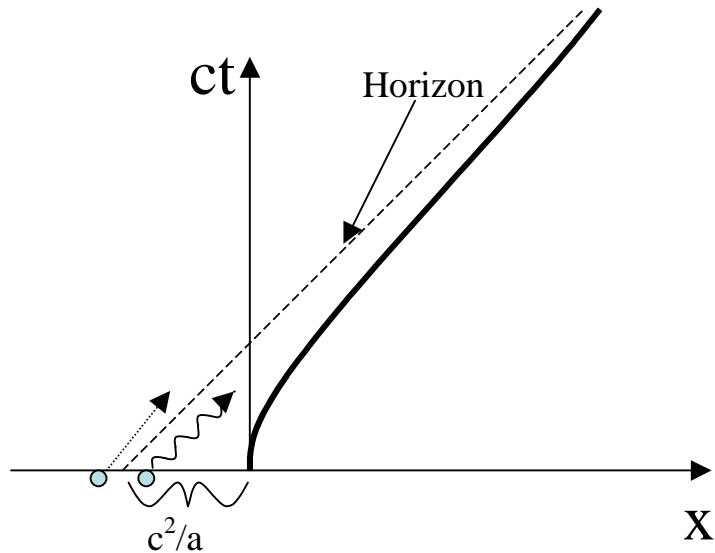

Horizon for accelerating observerObviously, nothing physically unusual happens at the horizon in Minkowski space as far as inertial observers can tell.  Yet, in neither coordinate system can a communication from left of the horizon reach the rocket.  This means that the region to the left of the horizon is causally disconnected from the rocket.

Unruh[4] demonstrated that the particle vacuum for the inertial Minkowski observers has components with radiation present for accelerating observers.  This radiation is associated with the information locally quantum correlated across the horizon lost to the accelerating observer.  The diagram in Figure 5demonstrates lost quantum coherence information associated with Figure 4.

Lost information about space-like correlations xamine the differential time for a photon to travel a differential distance dx from the initial position x $c\, d\tau_x^{from\,rear} = \dfrac{dL}{1 - \dfrac{aL}{c^2}} = \dfrac{d\tilde{x}}{1 - \dfrac{a(\tilde{x}_{observer} - \tilde{x})}{c^2}}$

If one chooses the horizon to correspond to position $\tilde{x}_{horizon} = 0$, this means that the observer coordinate is given by $\tilde{x}_{observer} = c^2/a$. This then cancels the 1 in the denominator above, giving a time interval associated with a photon $c\, d\tau_x^{from\,rear} = \dfrac{d\tilde{x}}{\left(\dfrac{a\,\tilde{x}}{c^2}\right)}$.

One can use the coordinates to develop how a standard clock ticks in a standard manner. The clock will be constructed by placing a photon source and detector at the left end of the clock, and using a rigid rod that utilizes microscopic forces to maintain a mirror at a fixed proper distance L to the right of the source/detector. One tick of the clock will correspond to the emission, reflection, and absorption of the light pulse. The components of this standard clock have trajectories shown in Figure 6.

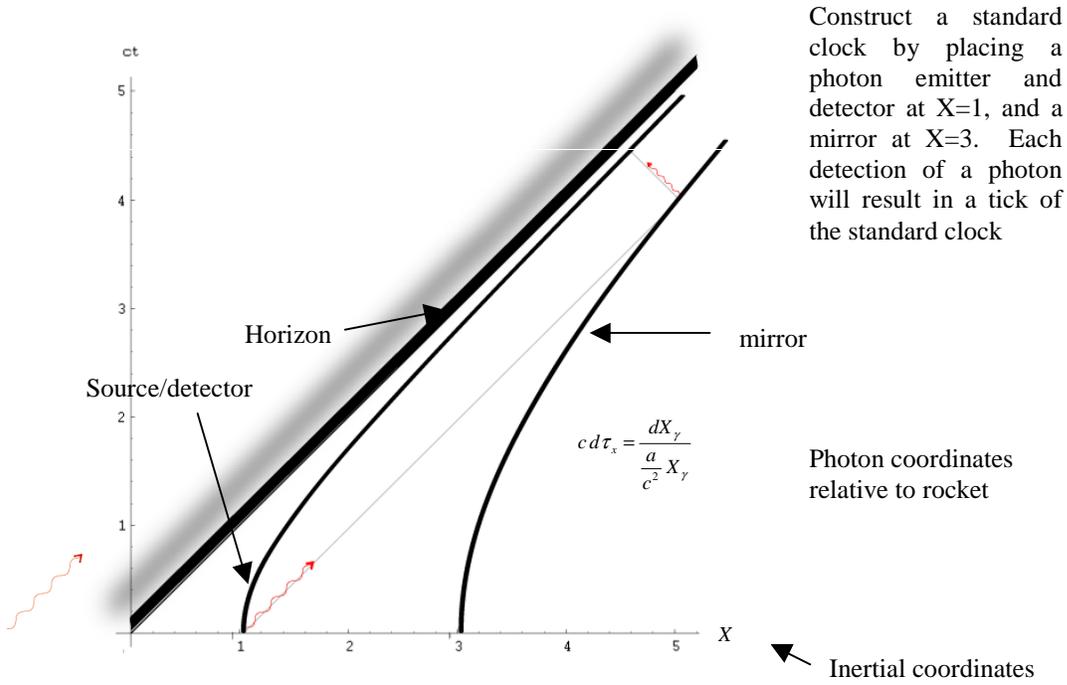

Construct a standard clock by placing a photon emitter and detector at X=1, and a mirror at X=3. Each detection of a photon will result in a tick of the standard clock

Photon coordinates relative to rocket

Inertial coordinates

Standard clock in Rindler space-time

As previously stated, the coordinate $\tilde{x}$ will be chosen to measure the proper distance from the horizon.. The photon coordinate from Figure 6must describe a null geodesic while maintaining $\tilde{x}$ as the proper distance from the horizon. The photon trajectory must satisfy the relationship established in Eq. (7). These properties are combined in the metric demonstrated in the following equation:

$$ds^2 = d\tilde{x}^2 \quad \text{when} \quad cd\tilde{t} = 0 = d\tilde{y} = d\tilde{z}$$

$$ds_\gamma^2 = 0 \Rightarrow 0 = -(c\,d\tilde{t})^2 + \left(\frac{d\tilde{x}_\gamma}{\frac{a}{c^2}\tilde{x}_\gamma}\right)^2$$

$$ds^2 = -\left(\frac{a}{c^2}\tilde{x}\right)^2 (c\,d\tilde{t})^2 + d\tilde{x}^2 + d\tilde{y}^2 + d\tilde{z}^2 = -c^2 d\tau^2$$

)The metric in Eq. (8) describes so called Rindler space-time, which has the following properties:

- A unique horizon at $\tilde{x} = 0$,
- Varying proper acceleration given by $a_{proper} = c^2 / \tilde{x}$.

Using the form of this metric at the source/detector end of the clock shown in Figure 6 parameterized by $\tilde{x}_o$ with a mirror located a distance $L$ to the right, the temporal interval between clock ticks is given by

$$\tau_f - \tau_o = \frac{a\tilde{x}_o}{c^2}\left(\tilde{t}_f - \tilde{t}_o\right) = 2\frac{\tilde{x}_o}{c}\log\left(1 + \frac{L}{\tilde{x}_o}\right).$$

)The clock is seen to operate well at any position away from $\tilde{x}_o = 0$.

The development of a locally accelerating standard clock provides a hint on the transition from special relativity to general relativity. The standard clock is seen to have the following properties:

- The size of the clock $L$ can (in principle) be made as small as desired. For small L the tick rate is simply that expected in flat space-time $\tilde{t}_f - \tilde{t}_o \cong 2L/c$, when $L << \tilde{x}_o$
- The rate at which a clock ticks clearly depends on its position $\tilde{x}_o$
- A key assumption of general relativity is the *principle of equivalence*, which can be expressed by asserting that the affects of being in an accelerating frame of reference is locally equivalent to a gravitational field in the opposite direction.

General relativity formalizes the general coordinate transformations that can incorporate global space-time characterizations consistent with the principle of equivalence.

## IV. Gravitation, General Relativity, and Quantum Mechanics

Gravitation is unique in that it is the only dynamics for which arbitrary test objects will undergo identical geodesic trajectories regardless of their dynamical coupling. The equivalence of the inertial mass from Newton's second law and the gravitational coupling constant makes gravity a unique dynamical field for the purpose of geometrodynamics, as seen in the equations

$$\vec{F} = m\frac{d^2\vec{r}}{dt^2} = m\vec{\nabla}\left(\frac{-GM}{r}\right) \quad \text{vs.} \quad m\frac{d^2\vec{r}}{dt^2} = q\vec{\nabla}\left(\frac{Q}{r}\right).$$

$$) ds^2 = d\xi^\mu \eta_{\mu\nu} d\xi^\nu, \text{ where } \eta = \begin{pmatrix} -1 & 0 & 0 & 0 \\ 0 & 1 & 0 & 0 \\ 0 & 0 & 1 & 0 \\ 0 & 0 & 0 & 1 \end{pmatrix}.$$

)The usual summation convention over repeated Greek indices is presumed in all equations that follow.

However, most often we utilize coordinates that are not freely falling, but rather are stationary relative to our observations. These general, curvilinear coordinates are related to the freely falling coordinates by the coordinate transformation x(ξ), or conversely, by the inverse coordinate transformation ξ(x). The unique curves defining trajectories in the inertial coordinates ξ are the straight lines associated with Newton's first law of motion. These curves map into the geodesics of the general curvilinear coordinates x. A straight line is characterized by having a vanishing second derivative with respect to the proper time associated with the particle trajectory

$$\frac{d^2\xi^\alpha}{d\tau^2} = 0.$$

)The trajectory can be likewise described using general curvilinear coordinates. The chain rule for partial derivatives defines the trajectory in these coordinates:

$$\frac{d^2 x^\beta}{d\tau^2} = \frac{d}{d\tau}\left(\frac{\partial x^\beta}{\partial \xi^\alpha}\frac{d\xi^\alpha}{d\tau}\right) = -\left(\frac{\partial x^\beta}{\partial \xi^\alpha}\frac{\partial^2 \xi^\alpha}{\partial x^\mu \partial x^\nu}\right)\frac{dx^\mu}{d\tau}\frac{dx^\nu}{d\tau} \equiv -\Gamma^\beta_{\mu\nu}\frac{dx^\mu}{d\tau}\frac{dx^\nu}{d\tau}.$$

)Alternatively, we can recall that a straight line is a unique curve on a space-time called a *geodesic*. A geodesic is the curve representing the shortest space-time distance between two points (or *events*) in the space-time. By extremizing the space-time distance

$$W = \int \sqrt{-\left(\frac{dx^\alpha}{d\tau}g_{\alpha\beta}(x)\frac{dx^\beta}{d\tau}\right)}\, d\tau$$

)the Geodesic Equation (13) is reproduced from the resultant Euler-Lagrange equations, with a form relating the connections $\Gamma$ of the affine space to the metric forms $g$ in the Riemannian space-time:

$$\Gamma^\beta_{\mu\nu} = \frac{1}{2}g^{\beta\alpha}\left\{\frac{\partial g_{\alpha\nu}}{\partial x^\mu} + \frac{\partial g_{\mu\alpha}}{\partial x^\nu} - \frac{\partial g_{\mu\nu}}{\partial x^\alpha}\right\}.$$

) A form for the metric that regenerates Newtonian gravitation for slow motions (v<<c) and weak fields ($\{GM/c^2 r\}$<<1) can be found using Newton's second law of motion:

$$\frac{d^2\vec{r}}{dt^2} = \vec{\nabla}\left(\frac{-GM}{r}\right)$$

$$\frac{d^2 r^j}{dt^2} \approx -\Gamma^j_{00}c^2 \cong -\frac{1}{2}\nabla^j g_{00} c^2$$

$$\Rightarrow g_{00} \cong -\left(1 - \frac{2GM}{c^2 r}\right)$$

)The final form in Eq. (16) is obtained by requiring that the metric form be that of flat Minkowski space-time far from the source of gravitation.

There is experimental evidence that quantum coherence is maintained by a static gravitational field. Experiments by Overhauser, et.al.[5] have demonstrated the gravitation of coherent neutrons diffracting from an apparatus whose orientation could be changed relative to the Earth's gravitational field. This means that (at least for stationary sources) gravitating systems maintain their quantum behavior. These experiments were also a test of the principle of equivalence (i.e., the motion of the observer does not break the coherence of an inertial system). Overhauser's apparatus is shown in Figure 7.

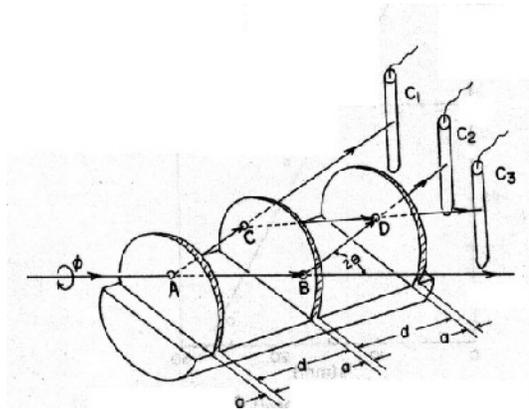

Overhauser's apparatus

The entire apparatus can be rotated about the axis AB. The detectors are indicated by C1, C2, and C3. The dimensions are given in Box (17)

$$\beta = (q_{grav} + q_{bend})\sin\varphi$$
$$q_{grav} = 4\pi\lambda\, g\, (2\pi\hbar)^{-2} m_n^2 \times$$
$$d(d + a\cos\theta)\tan\theta$$
$$\lambda = 1.445\, \overset{o}{A}\,,\; \theta = 22.1°\; Bragg\; angle$$
$$a = 0.2 cm\,,\; d = 3.5 cm$$
$$I_2 = \gamma - \alpha\cos\beta$$
$$I_3 = \alpha(1 + \cos\beta)$$
$$\frac{\gamma}{\alpha} = 2.6\,,\; \text{average intensity ratio}$$
$$q_{grav} + q_{bend} = 54.3\,,\; q_{grav} = 59.6$$

)

and the observed neutron counts are demonstrated in Figure 8.

$I_2 - I_3$

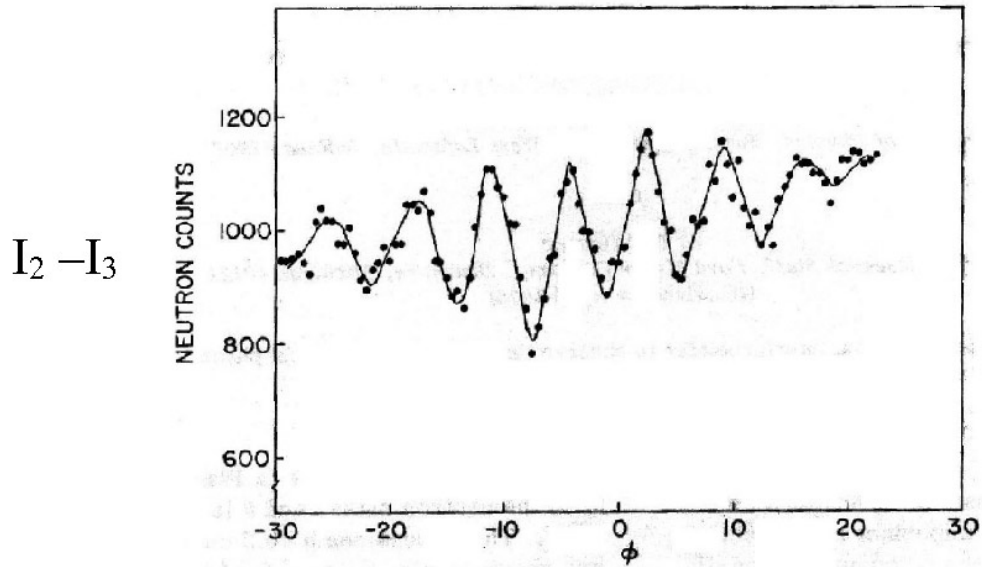

Gravitating coherent neutron diffraction data

The intensity is expected to vary based on the gravitational potential difference between paths through aperture B vs. C. Since the data clearly indicates that the coherence of the quantum neutron state is maintained, this implies that the action of gravitation did not localize the state of the neutron during its traversal from source to detection.

Therefore, one expects to be able to describe quantum systems in a gravitational field. How does one do quantum mechanics in a gravitating environment? The Einstein equation relates a quantity $G$ (the Einstein tensor) that is geometrically conserved due to the Bianchi identity (or the Jacobi identity for covariant derivatives) to the dynamically conserved energy-momentum tensor $T$:

$$G_{\mu\nu} \equiv R_{\mu\nu} - \frac{1}{2} g_{\mu\nu} R = -\frac{8\pi G}{c^4} T_{\mu\nu}$$

)

where the general form of the invariant metric is given by

$$ds^2 = g_{\mu\nu} dx^\mu dx^\nu.$$

)In general, a physical model for a quantum system is constructed from an invariant Lagrangian $\tilde{L}$ from which a scalar density form of the Lagrangian is constructed for the action that generates the equations of motion:

$$\text{Action} \quad W = \int L d^4 x \quad, \quad \text{scalar density } L = \sqrt{-g}\ \tilde{L}$$
$$\tilde{L} = \text{Minkowski microscopic physical model} \quad, \quad \eta \to g$$

)

The energy-momentum tensor can be obtained from variations of this action with respect to components of the metric tensor g

$$\delta W \equiv \frac{1}{2} \int d^4 x \sqrt{-g}\, T^{\mu\nu} \delta g_{\mu\nu} \quad, \quad T^{\mu\nu}{}_{;\nu} = 0.$$

)These steps outline the procedure by which one does quantum mechanics in a space-time background parameterized by metric $g$.

## V. Spherically Symmetric Black Holes

The geometries to be explored in this presentation will all be assumed to have spherical symmetry. The form of a general, static, spherically symmetric space-time metric is given by

$$ds^2 = -A(r)\,c^2 dt^2 + B(r)\,dr^2 + r^2 d\theta^2 + r^2 \sin^2\theta\, d\varphi^2.$$

)In the region exterior to the mass-energy source, the geometry satisfies the vacuum form of the Einstein equation:

$$G_{\mu\nu} = 0 \quad \Rightarrow \quad R_{\mu\nu} = 0$$

$$\frac{d}{dr}(A(r)B(r)) = 0$$

)This means that the coefficient of the temporal term in the metric is the inverse of the coefficient of the radial term. The vacuum metric then satisfies the form defining *Schwarzschild geometry*:

$$ds^2 = -\left(1 - \frac{2GM}{c^2 r}\right)c^2 dt^2 + \frac{dr^2}{\left(1 - \frac{2GM}{c^2 r}\right)} + r^2 d\theta^2 + r^2 \sin^2\theta\, d\varphi^2.$$

)The metric defines proper distance intervals for radial and angular displacements as follows:

$$dl_r \equiv d\rho = \frac{dr}{\sqrt{\left(1 - \frac{2GM}{c^2 r}\right)}} \quad \text{proper radial distance}$$

$$dl_\theta = r\,d\theta \qquad \text{proper azimuthal distance}$$

$$dl_\varphi = r\sin\theta\, d\varphi \qquad \text{proper polar distance}$$

$$A_R = 4\pi R^2 \qquad \text{Area of sphere of radius } R$$

)The Schwarzschild coordinates are seen to asymptotically ($r\to\infty$) correspond to those of Minkowski (flat) space-time. The Schwarzschild radial coordinate does not measure proper distance to the center of gravitation, but rather is the radial measure for angular displacements, as demonstrated in Eq. (25).

## Schwarzschild black holes

It is convenient to define the Schwarzschild radius $R_S \equiv \dfrac{2MG}{c^2}$, which has dimensions of length. The Schwarzschild metric can then be expressed by

$$ds^2 = -\left(1 - \frac{R_S}{r}\right)c^2 dt^2 + \frac{dr^2}{\left(1 - \dfrac{R_S}{r}\right)} + r^2 d\theta^2 + r^2 \sin^2\theta d\varphi^2$$

)

Several features of interest can be noted:

- If the vacuum solution holds down to $r=R_S$, then the geometry has a coordinate singularity.
- Physical tidal forces, etc. depend on curvature components, which all scale like $(1/R_S)^2$, and are therefore finite. Therefore this is ***not*** a physical singularity at $r=R_S$. However, the singularity at $r=0$ ***is*** physical.
- A freely falling observer from a radial coordinate $R$ will not reach the horizon in finite Schwarzschild time $t$, but will reach the singularity at $r=0$ in finite proper time $\tau = \dfrac{\pi}{2} R \left(\dfrac{R}{2MG}\right)^{1/2}$.

The large scale structure of Schwarzschild space-time will next be explored. In a manner analogous to that utilized in the development of finite conformal coordinates covering Minkowski space-time, the coordinates to be

utilized for the Schwarzschild Penrose diagram are as follows: Horizontal axis label (spatial)

$$Y_{\rightarrow} = \frac{-\tanh\left(\frac{ct - \left(r + R_S \log\left(\frac{r}{R_S} - 1\right)\right)}{R_S}\right) + \tanh\left(\frac{ct + \left(r + R_S \log\left(\frac{r}{R_S} - 1\right)\right)}{R_S}\right)}{\sqrt{2}}$$

Vertical axis label (temporal)

$$Y_{\uparrow} = \frac{\tanh\left(\frac{ct - \left(r + R_S \log\left(\frac{r}{R_S} - 1\right)\right)}{R_S}\right) + \tanh\left(\frac{ct + \left(r + R_S \log\left(\frac{r}{R_S} - 1\right)\right)}{R_S}\right)}{\sqrt{2}}$$

The resulting Schwarzschild Penrose diagram is demonstrated in Figure 9. The right causal regions inclusive of the future singularity is of relevance for describing a static, spherically symmetric black hole. Coordinates curves for the Schwarzschild temporal and radial coordinates have been drawn in the region exterior to the horizon. It is important to notice that the future singularity corresponding to the center of gravitation *r=0* is a space-like line (temporally stationary parameterized by spatial displacements) for a black hole, while *r=0* was a time-like line (stationary in space undergoing intervals of temporal displacement) in Minkowski space-time. Since the geometry is static, such a black hole is eternal. However, it is of interest to examine an example by which a black hole could be formed in a space-time initially devoid of any physical singularity. Such an example will next be explored.

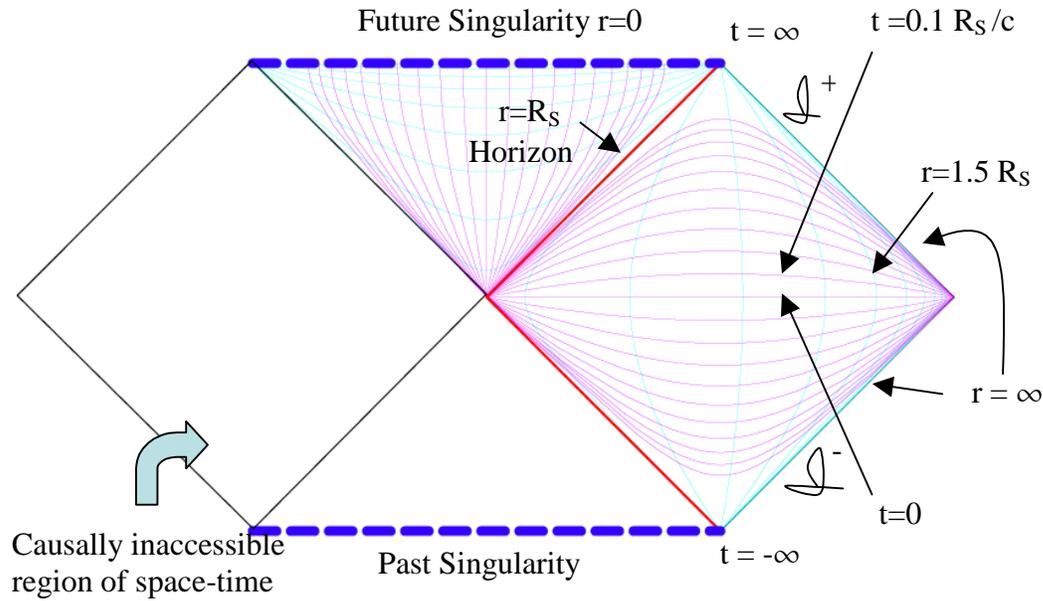

Schwarzschild Penrose diagram

**Creating a (Classical) Black Hole** Next, consider an in-falling spherical shell of light with a finite shell width. It will be assumed that the shell has total energy E. Eventually, this shell will cross its own Schwarzschild radius defined by $R_S=2\,G\,E/c^4$. By Birkoff's theorem for this spherically symmetric geometry, there is insignificant gravitation in regions interior to the incoming symmetric energy shell, whereas test objects in the region external to the shell gravitate as if the shell is a source located at the center of gravitation. However, eventually this shell will reach the center, creating a physical singularity at $r=0$. Once the singularity forms, the vertical time-like curve $r=0$ representing the center in Minkowski (negligible curvatures) space-time becomes the horizontal physical singularity $r=0$ representing the center in Schwarzschild (high curvatures) space-time. This means that there is the possibility that some outgoing photons could be emitted from the origin $r=0$ that would eventually hit the singularity $r=0$ at a future time. The *horizon* for this geometry is defined as the outermost surface of outgoing light-like trajectories that cannot reach spatial infinity. It turns out that the horizon actually forms prior to its crossing the in-falling energy shell. Therefore, horizons are defined in terms of the *global* geometry, not any local characteristics of the space-time.

How does one construct a Penrose diagram for the global space-time representing this situation? The region interior to the in-falling energy shell should be parameterized by the Minkowski space-time describing region A in Figure 10.

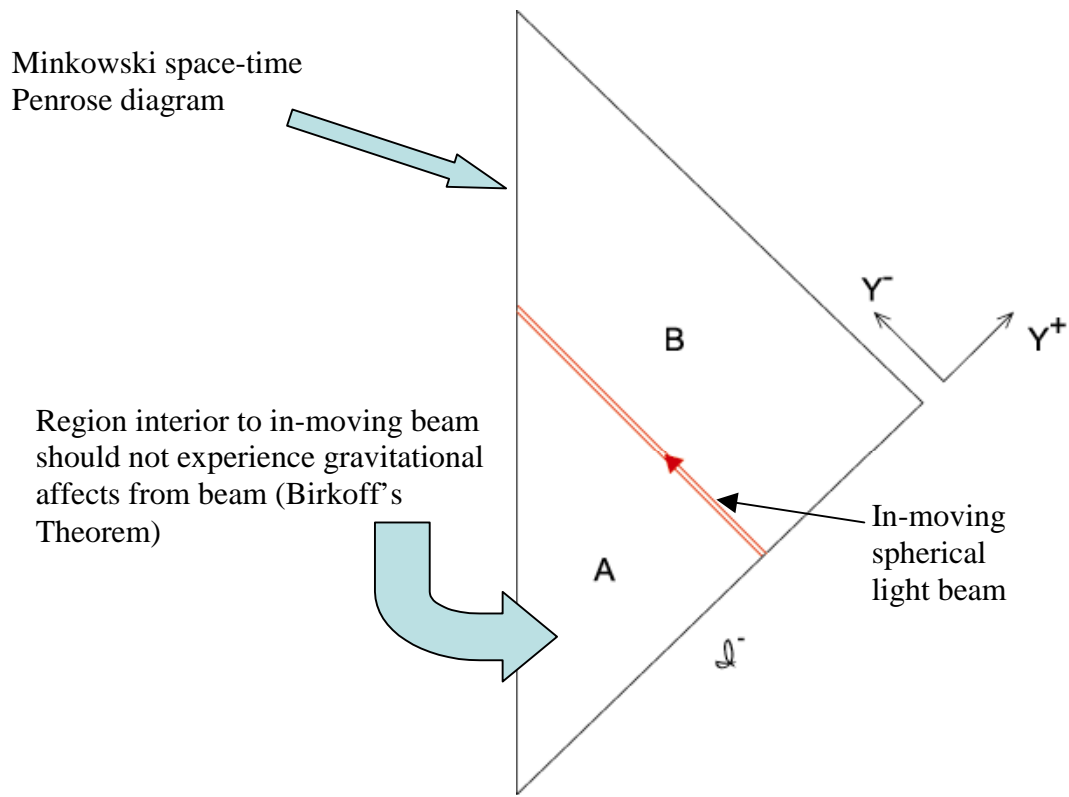

Interior Minkowski region Similarly, the region exterior to the in-falling energy shell should be parameterized using the Schwarzschild space-time describing region B' in Figure 11.

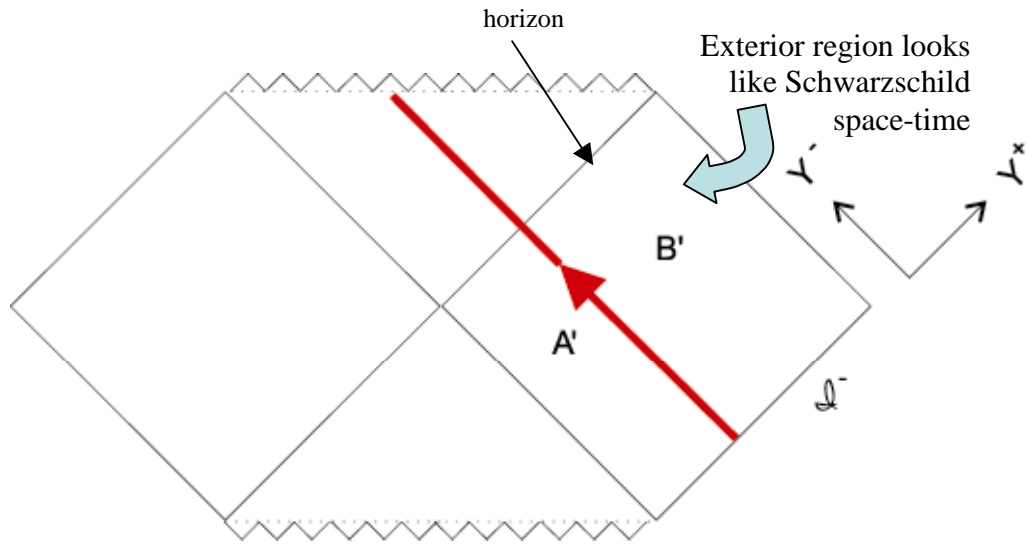

Exterior Schwarzshild region

The global Penrose diagram representing the formation and sustenance of this black hole is obtain by the appropriate joining of the regional space-time descriptions. This is done by "pasting" the causally adjacent regions associated with the separate regional *patches* in a manner that preserves the areas $A = 4\pi r^2$ associated with the suppressed $\theta$ and $\varphi$ coordinate symmetries, since angular displacements take the same form in both Minkowski and Schwarzschild geometries for a given radial coordinate *r*. This procedure is demonstrated in Figure 12.

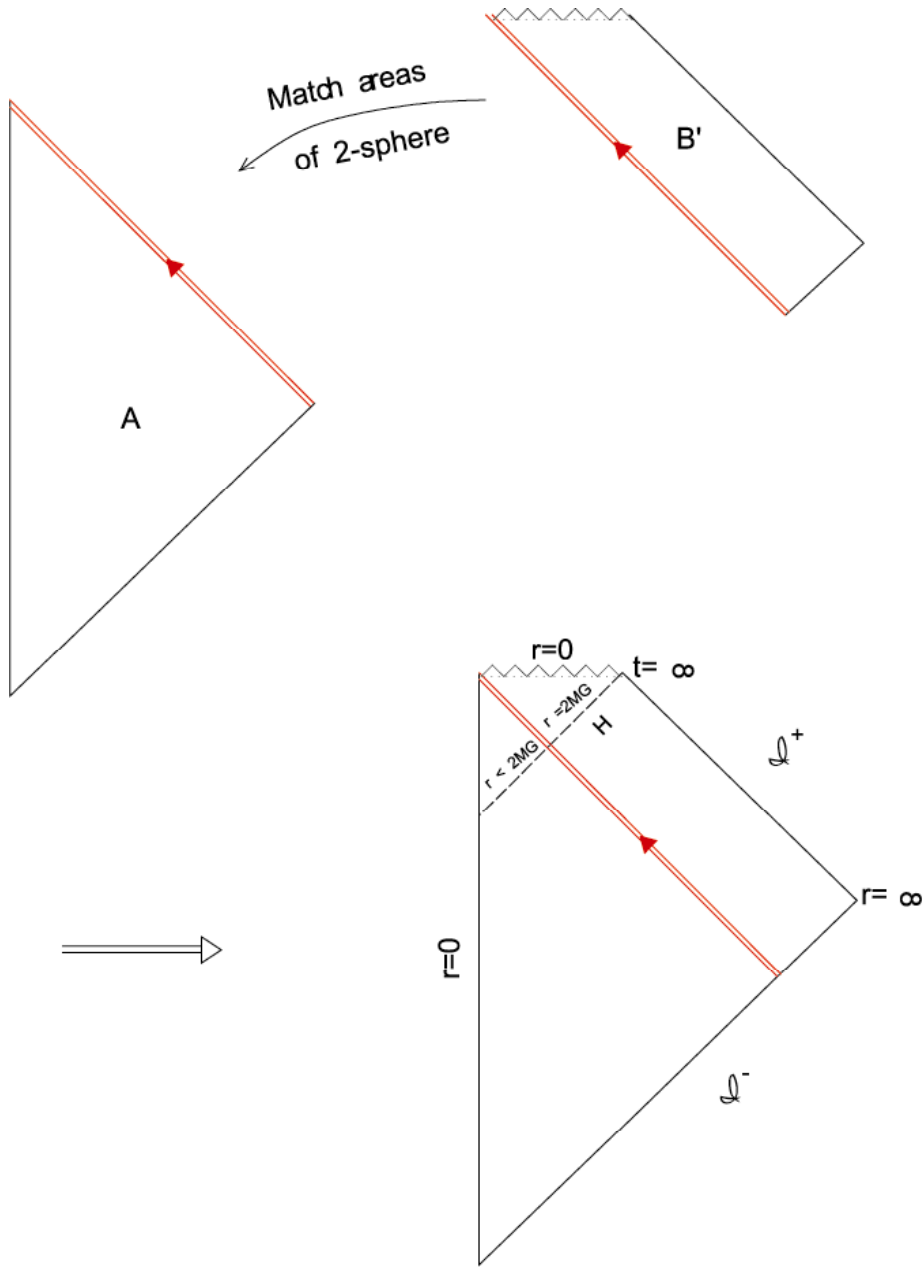

Joining of regional space-time descriptions

The resulting Penrose diagram is represented in Figure 13. As previously mentioned, a physical singularity forms once the in-falling shell reaches $r=0$. Any light-like surface has a slope of $45°$ on such a diagram, making causal relationships between regions directly observable.

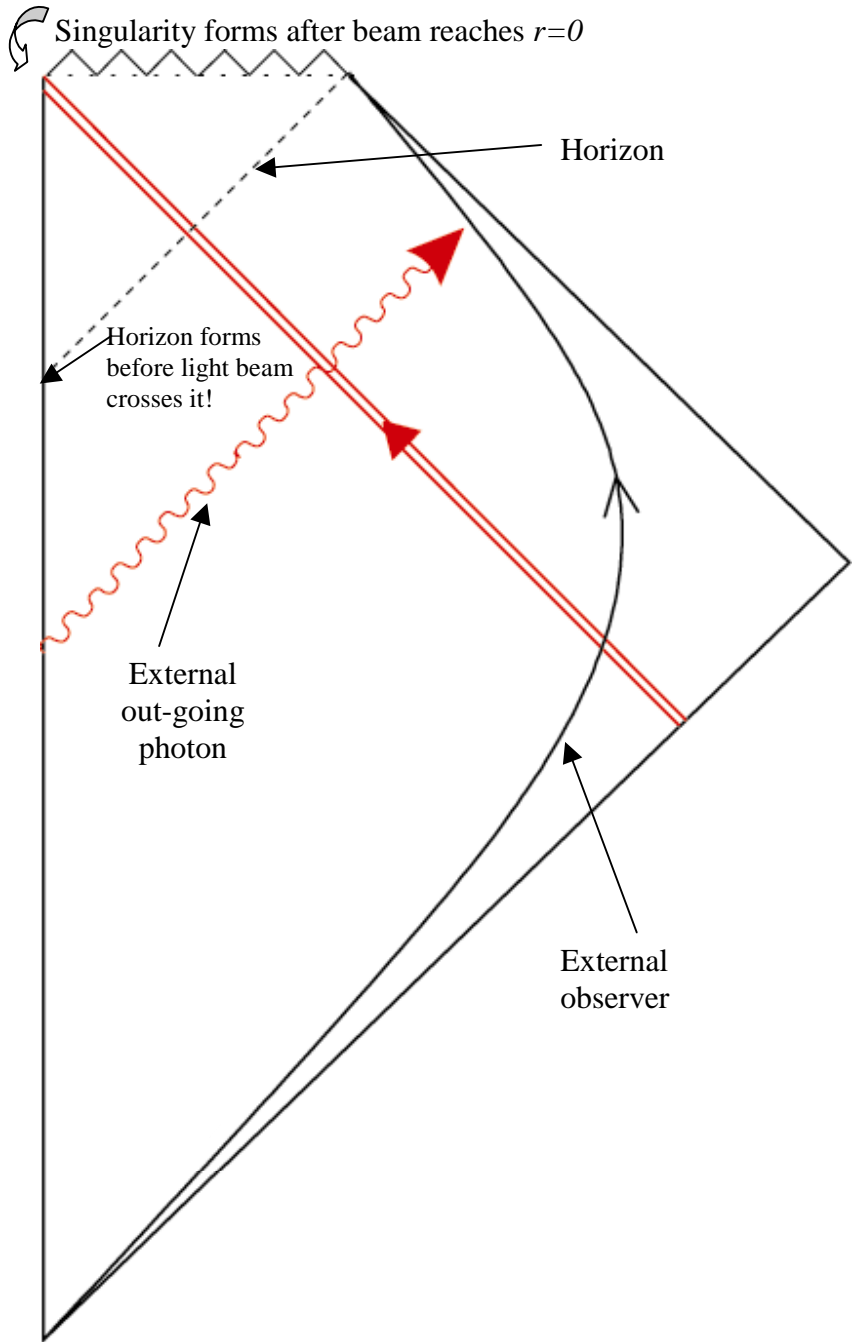

An evolved, non-evaporating black hole Any photon emitted from the region to the left of the horizon will eventually hit the singularity. This is true even if the photon were emitted during the low curvature period prior to the time that the in-falling energy shell crosses the photon trajectory. This exemplifies the global nature of the horizon.

## Near horizon Schwarzschild geometry

The behavior of the Schwarzschild space-time in the region very near the Schwarzschild radius will next be explored. The Schwarzschild metric will be parameterized in terms of the proper radial distance ρ: Very near the Schwarzschild radius $r \approx R_S = 2GM/c^2$, one can replace non-singular radial coordinates by the Schwarzschild radius:

$$d\rho = \frac{dr}{\sqrt{1 - \frac{R_S}{r}}} \approx \sqrt{R_S} \frac{dr}{\sqrt{r - R_S}}.$$

)The integration in Eq. (29) is straightforward, yielding the proper distance to the horizon given by

$$\rho \equiv 2\sqrt{R_S(r - R_S)}$$

)The Schwarzschild metric is then approximated by

$$ds^2 \approx -\left(1 - \frac{R_S}{r}\right)c^2 dt^2 + d\rho^2 + r^2 d\theta^2 + r^2 \sin^2\theta d\varphi^2$$

$$\approx -\rho^2 \frac{c^2 dt^2}{4R_S^2} + d\rho^2 + r^2 d\theta^2 + r^2 \sin^2\theta d\varphi^2$$

)It should not be too surprising that the near-horizon Schwarzschild coordinates are related to the Rindler space-time explored for a constantly accelerating observer in Minkowski space-time. Very near the horizon, the tangential coordinates behave similarly to how tangential spherical coordinates behave near the surface of the Earth, leading some to have proclaimed the Earth flat in the past. The acceleration $a = \dfrac{c^2}{2R_S}$ that would be associated with this Rindler space-time will be directly related to the temperature of the horizon (this is not the same as the proper acceleration of the Rindler space time, which is independent of $a$).

The coordinate singularity associated with the Schwarzschild radius can be explored by making the following identifications:

$$\omega \equiv \frac{ct}{2R_S} = \frac{c^3}{4GM} t \quad \text{dimensionless Rindler time}$$

$$a_{proper} = \frac{c^2}{\rho} \quad \text{proper acceleration}$$

$$ds^2 \approx -\rho^2 d\omega^2 + d\rho^2 + |d\bar{x}_\perp|^2 \quad \text{Rindler metric}$$

)The coordinate singularity at $\rho=0$ is seen to be the hyperbolic analog of the polar coordinate singularity representing angular ambiguity at $r=0$. Also, the Schwarzschild time scales as if there were an asymptotic acceleration

$$\frac{a_\infty}{c^2} \rho\, cdt = \rho\, d\omega = \frac{c^2}{4GM} \rho\, cdt \quad \rightarrow \quad a_\infty \sim \frac{c^4}{4GM}.$$

)Since there is no asymptotic proper acceleration, this temporal rescaling results in an asymptotic temperature that the distant Schwarzschild observer associates with the black hole. This will be demonstrated in the next sub-section. However, the details briefly explained in the following sub-section are not necessary for the results of the next chapter, and can be skipped if the justifications for black hole evaporation are not of interest to the reader.

## Entropy and temperature

A fundamental characteristic of a black hole is its finite light-like surface defining its horizon. The existence of any horizon implies an information deficit in the region causally excluded by the horizon. Complete information from quantum correlations across the horizon cannot be transmitted from the excluded region. Several quantum states beyond the horizon will therefore correspond to a given state in the causally accessible region. These states must be handled statistically with regards to the physics describing the accessible region. Statistical physics assigns an entropy associated with the number of microscopic configurations that can correspond to a given "course" measurement. This entropy describes the disordered internal energy that parameterizes the *heat* in the first law of thermodynamics, and the non-decreasing degree of *randomness* in the second law of thermodynamics. The classical thermodynamic parameters are obtained by statistically averaging over the possible incoherent configurations associated with a given course measurement. The partition function $Z$ is the statistical factor that normalizes the probability distributions. The thermodynamic energy, free energy, temperature, entropy, and partition function are related through the representation independent relationship given by $<\hat{H}> = F - T\frac{\partial F}{\partial T} = F + TS$ , $Z = e^{-\beta F} = Tre^{-\beta \hat{H}}$

)where $\beta = \frac{1}{k_B T}$. The thermal density matrix operator, which is the statistical measure of the relative distribution of the incoherent states, is given by

$$\hat{\rho} = \frac{e^{-\beta\hat{H}}}{Z} = e^{-\beta(\hat{H}-F)}$$

The statistical weights are properly normalized by the condition $Tr\bar{\rho} = 1$. Classical thermodynamic parameters (like pressure, energy, etc.) are obtained by thermally averaging using weights determined by the density matrix

(36)o evaluate the trace in Eq. (34), one can insert a complete set of energy basis states, obtaining

$$Z \propto \int dE e^{-\beta E + S(E)/k_B}, \quad \text{alternatively} \quad Z = \int dE \eta(E) e^{-\beta E}$$

where the density of states factor is related to the number of microstates associated with a given energy configuration $\eta(E) \propto e^{S(E)/k_B}$.

A brief derivation of the temperature associated with the horizon can be given using straightforward techniques from statistical physics. An arbitrary thermal representation operator $<\hat{Q}(\theta)>_{thermal} \equiv Tr\,\hat{\rho}\hat{Q}(\theta)$

(38)yclicity of the trace and the form of the density operator in Eq. (35), thermal averages of thermal representation operators can be shown to satisfy a periodicity relationship $\langle\hat{Q}(\theta)\rangle = Tr\hat{\rho}Q(\theta) = \langle\hat{Q}(\theta+\beta)\rangle$. Therefore, thermal representation operators are periodic with period $\theta : 0 \to \beta$.

- Euclidean metric calculation of temperature

    Consider Rindler space from Eq. (32), with the metric

$$ds^2 = -\rho^2 d\omega^2 + d\rho^2 + dx_\perp^2$$

where $dx_\perp$ represents the tangent plane to the black hole horizon and $\omega$ is the (dimensionless) Rindler time. Similarly to the procedure described for Minkowski space-time, thermodynamics, can be formally related to quantum dynamics in Rindler space-time by transforming to the Euclidean form of the metric using $-i\omega \to \theta_R$:

$$ds^2_{Euclidean} = \rho^2 \, d\theta_R^2 + d\rho^2 + dx_\perp^2$$

) The *conical* angle $\theta_R$ is seen to be periodic, varying from 0 to $2\pi$. This *inverse temperature* is the Euclidean "rotation angle" around the $x_\perp$ axes. However, we previously demonstrated that the Schwarzschild time is related to the Rindler time by $\omega = \dfrac{ct}{2R_S}$. This relates the Euclidean form of the Schwarzschild time $t \to i\hbar\theta$ to that of Rindler time in the form

$$\theta_R = \frac{\hbar c \theta}{2R_S}.$$

)The Rindler Euclidean form is periodic in $\theta_R$ with period $2\pi$, while the Schwarzschild Euclidean form is periodic in $\theta$ with period $\beta = \dfrac{1}{k_B T}$. Equation (41) then gives the most direct calculation of the Hawking temperature that a distant Schwarzschild observer associates with the horizon:

$$2\pi = \frac{\hbar c \beta}{2R_s} \quad \Rightarrow \quad k_B T_{Horizon} = \frac{\hbar c^3}{8\pi GM}$$

) Once a temperature has been established for the horizon, the conjugate entropy can be determined assuming that the first law of thermodynamics is satisfied by the black hole. This law relates the mechanical equivalence of heat, and in the absence of external work done by expansion against empty space, it takes the form

$$dE = dMc^2 = T \, dS.$$

)A direct substitution of *T* from Eq. (42) results in the form of the entropy of the black hole:

$$S = \frac{4\pi GM^2}{\hbar c}k_B + const \quad \Rightarrow \quad S = \frac{k_B c^3}{\hbar}\frac{4\pi R_S^2}{4G} = \frac{k_B c^3}{\hbar}\frac{Area}{4G}.$$

)For a black hole, the quantum states of the region interior to the horizon that correspond to a given external measurement should be statistically summed over[1]. Therefore, expectation values of any quantum operator in the exterior region can be expressed in traces using the microcanonical density matrix form $\rho_{exterior}(b,b') = \sum_a \psi^*(a,b)\psi(a,b')$, where the interior quantum states are labeled a.

- Thermal Path Integrals

An alternative derivation of the entropy that directly utilizes the form of the Einstein action to be inserted into the form of the partition function will be briefly sketched. The partition function is directly related to the vacuum functional in the path integral formulation using an Euclidean extension for the time parameter. The Lagrangian form that generates the Einstein equation is directly proportional to the Ricci scalar:

$$W_{Euclidean} = \frac{1}{16\pi G}\int \sqrt{g}\, R_{Euclidean}\, d^4x \quad \Rightarrow \quad Z = e^{-\theta_R F} = e^{-W_{Euclidean}/\hbar}$$

)The conical singularity corresponding to ρ=0 is degenerate in the Euclidean Rindler time parameter $\theta_R = i\omega$. Thermodynamic parameters are determined by derivatives with respect to $\theta_R$. However, since the Rindler Euclidean parameter takes on the constant value 2π, it is difficult to take such derivatives. To calculate the entropy one uses the conical deficit angle defined by modifying the range of the Euclidean parameter $0 \leq \theta_R \leq 2\pi - \varepsilon$. The singular nature of the cone for

$\rho \to 0$ can be obtained by parameterizing the eccentricity of a hyperboloid of revolution and taking the appropriate limit. It is convenient to define the covering parameter $\lambda \equiv 1 - \dfrac{\varepsilon}{2\pi}$ to express the desired hyperboloid in terms of the deficit angle $\varepsilon$.

$$\vec{r} = \left( \lambda \rho \cos \frac{\theta_R}{\lambda}, \lambda \rho \sin \frac{\theta_R}{\lambda}, \frac{\sqrt{1-\lambda^2}}{\lambda}, \sqrt{\lambda^2 \rho^2 + \delta} \right)$$

)This 2-surface becomes the cone corresponding to Euclidean Rindler space when $\delta \to 0$, and has the desired deficit in $\theta_R : 0 \to 2\pi - \varepsilon$. The metric then takes the modified form

$$ds^2 = \rho^2 d\theta_R{}^2 + \frac{\lambda^2(\rho^2 + \delta)}{\lambda^2 \rho^2 + \delta} d\rho^2 + dx_\perp^2$$

)A straightforward calculation gives the determinant of the metric and the Euclidean curvature scalar:

$$\sqrt{g} = \lambda^2 \rho \sqrt{\frac{\rho^2 + \delta}{\lambda^2 \rho^2 + \delta}}$$

$$R_{Euclidean} = \frac{2\delta(1-\lambda^2)}{\lambda^2(\rho^2 + \delta)^2}$$

)One can see that the curvature scalar becomes singular for $\rho \to 0$ for the cone ($\delta \to 0$), and vanishes elsewhere. Finally, the form of the action can be evaluated

$$W_{Euclidean} = \frac{1}{16\pi G} 2\pi \, Area \left[ -\frac{\sqrt{\lambda^2 \rho^2 + \delta}}{\sqrt{\rho^2 + \delta}} \right]_0^\infty = \frac{1}{16\pi G}(2\pi) Area \, 2(1-\lambda) = \frac{A\varepsilon}{8\pi G}.$$

)Thus, the entropy is given by

$$S = -\theta_R^2 \frac{\partial}{\partial \theta_R}\left(\frac{\log Z}{\theta_R}\right) = -\theta_R^2 \frac{\partial}{\partial \theta_R}\left(\frac{(2\pi - \theta_R)}{\theta_R}\frac{A}{8\pi G}\right)\bigg|_{\theta_R = 2\pi} = \frac{k_B c^3}{\hbar}\frac{A}{4G}$$

)It is of interest to note that the entropy is of order $\frac{1}{\hbar}$, and has no obvious interpretation as to the form of any microscopic states that are counted to generate this entropy. In particular, it means that a black hole has infinite entropy as $\hbar \to 0$, implying that its entropy is non-perturbative with respect to it's quantum nature.

Returning again to the form of the Hawking temperature $T_{Hawking} = \frac{\hbar c^3}{8\pi G M k_B}$ and the entropy $S = \frac{k_B c^3}{\hbar}\frac{A}{4G}$, the following point are of interest for the discussion:

- Entropy is proportional to the *Area* of the horizon, **not** any *volume*. This is the basis of holography, since information is extensive on a surface rather than the bulk volume.
- For a freely falling observer, no horizon (no persistent causually inaccessible region of space) means **no** temperature without violating the principle of equivalence
- Entropy cannot be perturbatively calculated in Planck's constant
- Finite temperature implies thermal radiations associated with the horizon and eventual evaporation
- Puzzle:    Is Asymptopia ($r = \infty$) static or freely falling?
- Evaporation of Black Holes

The expected existence of radiations associated with the thermal nature of the horizon of a black hole implies that its mass/energy slowly evaporates away. The evaporation rate will be sensitive to details of microscopic physics, like how many low mass particle species have energies comparable to the thermal energies of a horizon of a given temperature. One can show[1] that primarily low angular momentum quanta escape (predominantly only s-waves). If one assumes s-wave quanta as 1+1 dimensional quantum fields at temperature $T_{Rindler}=1/2\pi$, and that the barrier height is comparable to the thermal energy $T_{Rindler}$, one concludes that approximately 1 quantum per unit Rindler time will escape. In terms of Schwarzschild time, this means that the flux will be about

$$flux \sim \frac{c^3}{MG}$$

$$energy/quantum \sim k_B T_{Hawking} = \frac{\hbar c^3}{8\pi MG}$$

)This predicts a luminosity given by $Luminosity = \frac{const}{(MG)^2} = -\frac{dM}{dt}$. Therefore, the evaporation time is seen to be of order $\tau_{evaporation} \sim \frac{M^3 G^2}{\hbar c^6}$. The candidate black holes at the centers of many galaxies are expected to have lifetimes much longer than the time since the initiation of the big bang. However, because of the Hawking radiation, black holes are not so black. The physics of a radiating star is very different from that of black hole. The typical wavelengths of stellar radiation are comparable to those of visible light. However, the typical wavelengths of black hole radiation is of the order of the thermal wavelength, which is of the order of the radius of the horizon. This means that images of the horizon formed from these radiations will always be fuzzy.

VI.    Temporally Dynamic Black Holes

If there can be accretion and evaporation, then the mass of a black hole, and all relevant scales, must assume a temporal dependency. An introduction of a naïve Schwarzschild temporal dependency for the mass indicates a new physical singularity at the Schwarzschild radius. Introducing a Schwarzschild time dependency to the mass of the black hole, one then calculate invariants associated with the geometric and energy content of the space-time:
$$R_M(t_S) \equiv \frac{2GM(t_S)}{c^2}$$
$$ds^2 = -\left(1 - \frac{R_M(t_S)}{r}\right)c^2 dt_S^2 + \frac{dr^2}{\left(1 - \frac{R_M(t_S)}{r}\right)} + r^2 d\theta^2 + r^2 \sin^2\theta d\varphi^2$$
$$\mathfrak{R} = -\frac{r\left(2\dot{R}_M^2 + (r - R_M)\ddot{R}_M\right)}{2(r - R_M)^3} = \frac{8\pi G}{c^4} g^{\mu\nu} T_{\mu\nu}$$
)where dots indicate derivatives with respect to the Schwarzschild time coordinate $ct_S$. One sees that the Ricci curvature scalar is singular at the coordinate anomaly $r=R_M$ for non-vanishing $\dot{R}_M \neq 0$. The Ricci scalar is seen from Eq. (52) to be directly related to invariant physical content using Einstein's equation
$$\mathfrak{R}_{\mu\nu} - \frac{1}{2} g_{\mu\nu} \mathfrak{R} = -\frac{8\pi G}{c^4} T_{\mu\nu} \quad \Rightarrow \quad \mathfrak{R} = \frac{8\pi G}{c^4} g^{\mu\nu} T_{\mu\nu}.$$
)The Ricci scalar should be non-singular at $r = R_M$ if this coordinate is only a coordinate anomaly. If the invariant Ricci scalar is singular, then the trace of the mixed energy-momentum tensor must likewise be singular. The singular behavior of the Ricci scalar represents a singularity in the local space-time that must be reflected in the physical content independent of the particular coordinate description. Such a singular physical structure would also be observed by a freely falling observer attempting to traverse the coordinate anomaly of the static observer. This violates what is expected by complementarity and the principle of equivalence.

## The River Model

The river model of a black hole utilizes a non-orthogonal river time $t_R$ to parameterize the temporal evolution of the spacetime. The spherically symmetric metric takes the form

$$ds^2 = -c^2 dt_R^2 + [dr - \beta(r)c\, dt_R]^2 + r^2 d\theta^2 + r^2 \sin^2\theta\, d\varphi^2$$

)where β(r) represents the "velocity of the flow" of the space-time river. The river time $t_R$ can be transformed into a diagonal time coordinate $t_*$ which transforms the metric into a Schwarzschild/deSitter metric form:

$$ct_R = ct_* - \int_{r_*}^{r} \frac{\beta(r')}{1-\beta^2(r')} dr'$$

)The diagonal form of the metric is then

$$ds^2 = -(1-\beta^2(r))c^2 dt_*^2 + \frac{dr^2}{(1-\beta^2(r))} + r^2 d\theta^2 + r^2 \sin^2\theta\, d\varphi^2$$

)indicating luminal river speed at the coordinate anomaly. A Schwarzschild black hole corresponds to a local river speed of $\beta(r) = \sqrt{\dfrac{R_S}{r}}$. It is important to note that for a Schwarzschild black hole, a radial parameter of coordinate correspondence $r_o$ that takes a value of infinity gives an infinite shift in coordinate times in Eq. (55). This suggests that the spatial Asymptopias (r=∞) of Schwarzschild and River black holes do not have overlapping temporal maps. In what follows, it will be assumed that the coordinates of correspondence between coordinate representations will be finite.

### Non-orthogonal coordinates

Inspired by the river model previously discussed, a temporal dependency will be introduced into the mass scale defining the black hole using the non-orthogonal temporal parameter $t_R$:

$$ds^2 = -c^2 dt_R^2 + \left[ dr + \sqrt{\frac{R_M(ct_R)}{r}} c dt_R \right]^2 + r^2 d\theta^2 + r^2 \sin^2\theta d\varphi^2.$$

)As the radial coordinate $r$ goes to infinity, the metric asymptotically takes the form of Minkowski space, defining the time $t_R$ as a global temporal parameter in the same manner as the Schwarzschild time $t_S$. A calculation of curvature invariants using this metric demonstrates that the Ricci scalar is finite everywhere away from a physical singularity at $r=0$.

$$\mathfrak{R} = -\frac{3\dot{R}_M}{2r^2}\sqrt{\frac{r}{R_M}} = \frac{8\pi G}{c^4} g^{\mu\nu} T_{\mu\nu}$$

)where dots in all subsequent equations represent derivatives with respect to the non-orthogonal temporal coordinate $ct_R$. Likewise energy content invariants are finite everywhere away from $r=0$, satisfying expectations that a freely-falling observer should be able to fall past a non-singular horizon.

The proper acceleration of a stationary observer remains singular at coordinate anomaly $r=R_M$

$$a_{proper} = \frac{R_M c^2}{2r^2}\sqrt{\frac{r}{r-R_M}}\left(1 + \frac{r^2 \dot{R}_M}{(r-R_M)R_M}\left(\frac{r}{R_M}\right)^{1/2}\right).$$

)However, unlike the static Schwarzschild case, there is a radial coordinate for which a stationary observer will experience a vanishing proper acceleration

$$r_\# \approx \frac{R_M}{1+\dot{R}_M}.$$

)

### Horizon and Mass Scale Evolution

A black hole's horizon is defined by its outgoing radial light-like (null) geodesic

$$ds_\gamma^2 = 0 = -c^2 dt_R^2 + \left[dr_\gamma + \sqrt{\frac{R_M(ct_R)}{r_\gamma}}\,c\,dt_R\right]^2 + r_\gamma^2 d\theta_\gamma^2 + r_\gamma^2 \sin^2\theta_\gamma d\varphi_\gamma^2$$

$$d\theta_\gamma = 0 = d\varphi_\gamma$$

)The dynamics of radially moving photons is therefore described by

$$\frac{dr_\gamma}{dct_R} = -\sqrt{\frac{R_M}{r_\gamma}} \pm 1$$

)where $R_M$ is the radial mass scale. Therefore, the horizon $R_H$ must satisfy

$$\frac{dR_H}{dct_R} = -\sqrt{\frac{R_M}{R_H}} + 1 \quad , \quad R_H = \frac{R_M}{(1-\dot{R}_H)^2}$$

)

This equation describes a relationship between the temporal dynamics of the horizon and that of the coordinate anomaly. One notes the following features:

- The horizon consists of the outermost set of null geodesics that cannot reach light-like future infinity $\mathcal{I}^+$. •Unlike static Schwarzschild geometry, $R_H \neq R_M$.

- Radially traveling photons located at $r=R_M$ are momentarily stationary, while a dynamic horizon is not!
- Radially outgoing photons between $R_H<r_\gamma<R_M$ will still escape the singularity if the black hole is evaporating!

### Accretion and evaporation

The expected form of a Penrose diagram representing the classical birth and death of a black hole[6] is given in Figure 14.

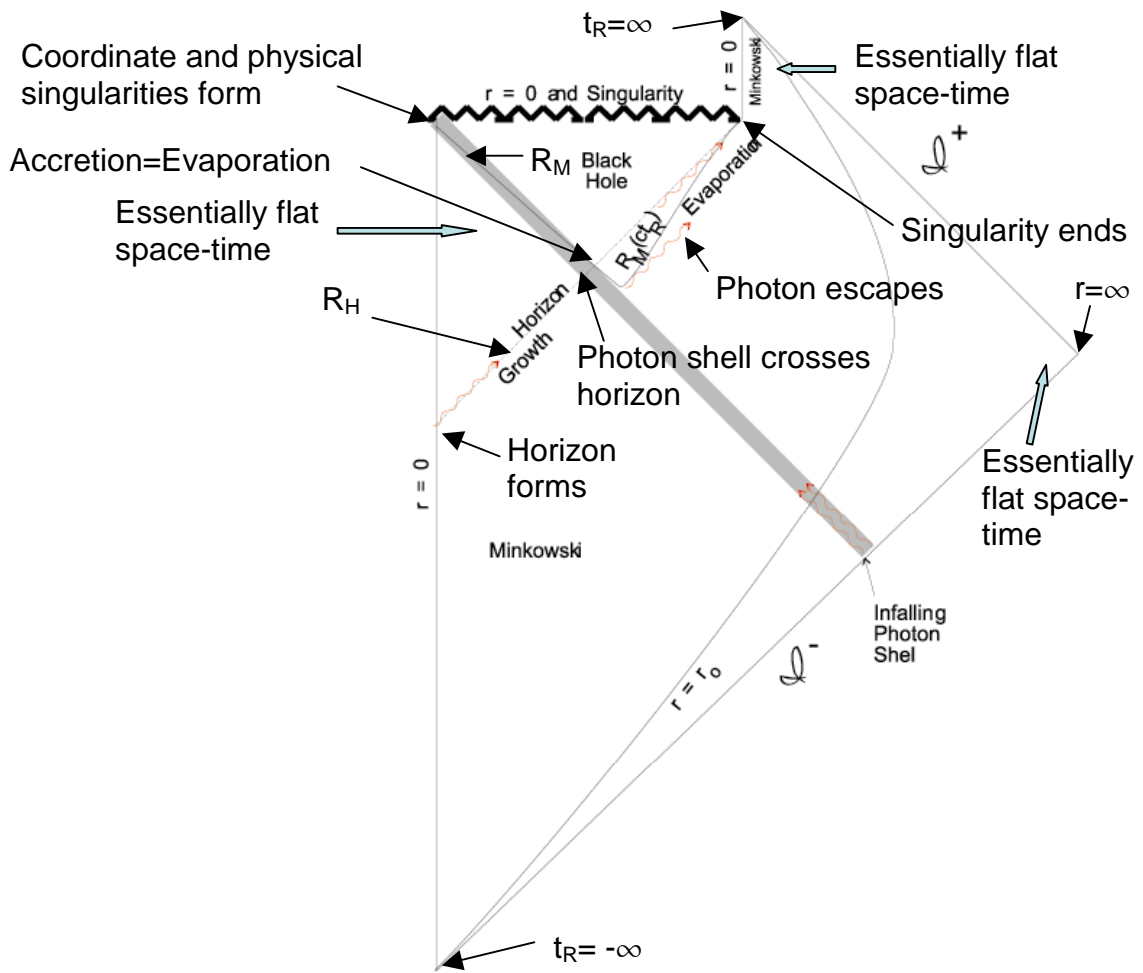

Figure 14 Penrose diagram for formation and evaporation of a black hole

In the Penrose diagram, an in-falling spherically symmetric photon shell contains an energy that manifests as the mass scale $M_o c^2$ of the space-time. This is the maximum mass contained in a spherical surface including any evaporants. The region interior to this photon shell is essentially flat because of Birkoff's theorem for spherically symmetric geometries, while the exterior region satisfies the geometry associated with a spherically symmetric mass distribution until that mass evaporates away. All elements of the in-falling shell will eventually cross the surface defining the horizon that bounds the region for which any outgoing light would eventually hit a singularity at *r=0*.

The Penrose diagram in Figure 14 joins an initially flat (Minkowski) geometry with a radially infalling photon shell along the causal boundary given by the leading surface of the photon shell. The thick band in the diagram originating at light-like past infinity $\mathcal{I}^-$ represents that photon shell, and the region beneath that band (interior to the shell) has negligible curvatures due to Birkoff's theorem. This lower triangular region is bounded on the left by the time-like curve representing *r=0*. Since the photon shell eventually reaches *r=0* forming a physical singularity (indicated by the jagged horizontal line on the diagram), there is a light-like surface representing the outermost set of out-going photons that will eventually hit the singularity that forms. This horizon is seen to be globally defined, having a non-vanishing radial coordinate $R_H > 0$ prior to the space-time point(s) when the in-falling photon shell crosses this horizon. However, the radial mass scale $R_M$ associated with the coordinate anomaly in the highly curved metric of the black hole geometry is seen to increase from a vanishing value to that appropriate to a Schwarzschild-like space-time as the photons in the shell cross this growing coordinate. As elements of the photon shell reach *r=0*, the curve *r=0* interior to the coordinate anomaly $R_M$ becomes the space-like singularity of increasing mass represented by the initiation of the horizontal jagged curve. The width of the photon shell represents the duration of the period of growth in the radial mass scale $R_M$. Increases in the radial mass scale are associated with local in-falling shell photons as they cross growing outgoing light-like scales, any of which would have represented the global horizon were the growth to have stopped at that stage. The curve $R_M(ct_R)$ grows away from the physical singularity at *r=0* after the leading surface of the in-falling photon shell initiates this singularity. In the space-time region with significant curvatures, the curve *r=0* tracks a physical singularity with a non-vanishing mass scale. The expected difference between the curve tracking the radial mass scale and the horizon $R_H$ has been exaggerated for emphasis. This difference is determined by the relation for the light-like curve given in Eq. (63). The curve $R_M(ct_R)$ crosses the global horizon $R_H$ when $\dot{R}_H = 0$, which occurs when the rate of mass growth is comparable to that of mass loss due to radiation. If the energy influx rate were to exactly match the evaporation rate for an extended period, the geometry would be expected to represent an essentially static Schwarzschild space, however, the Penrose diagram must represent the large scale structure of the space-time. For the case being examined, a photon emitted from $R_M$ is able to escape hitting the singularity because of the shrinking of the mass scale due to evaporation. Since the radial mass scale is associated with the curved metric, radial coordinates associated with it are determined relative to the jagged singularity *r=0* (not the Minkowski-like *r=0*). During growth, the coordinate anomaly $R_M$ has a value less than the radial coordinate of the horizon, whereas during evaporation the horizon has radial coordinate less than the radial mass scale. The physical singularity *r=0* and the coordinate anomaly $R_M$ are seen to vanish together, leaving a (shifted) time-like curve *r=0* associated with the final low curvature Minkowski-like space-time, represented as the upper triangular region in the diagram subsequent to complete evaporation of the singularity internal to the surface of final radiations. Some aspects of the quantum mechanics that can be done in such a space-time has been discussed in reference [6]. The initiation and dissolution of the physical singularity likely involve significant quantum behaviors, and will not be discussed in this presentation,

VII.  Discussions and Conclusions

In conclusion, it is noted that a naïve introduction of temporal dependencies into black hole dynamics using the Schwarzschild time introduces new physically singular behavior at the coordinate anomaly associated with static Schwarzschild observers. Such black holes probably couldn't grow in a direct way, since any in-falling matter would encounter severe forces approaching the horizon. Complementarity asserts that no observer should witness a violation of any law of nature. An inertial observer should not detect the presence of an accelerating observer's horizon. Otherwise, the principle of equivalence requires complicated caveats associated with purely gravitational radiations.

Black holes can be described using alternative temporal formulations with differing asymptotic behaviors. The temporal dependency that has been examined in this presentation parameterizes an asymptotically orthogonal time corresponding to a flat Minkowski space that takes a non-orthogonal spatial component in the near region that cancels singular temporal behavior near the coordinate anomaly. The horizon of a dynamic black hole then does not coincide precisely with the coordinate anomaly, giving natural scales for a *stretched horizon*. The temporal dependency of a horizon and radial mass scale is found to qualitatively modify the local coordinate structure of the space-time. In-falling observers notice no unusual structure or energy content as they traverse the horizon of static observers. Information is thermalized due to the finite extent of a physical singularity at *r=0* dressed with a horizon of finite duration. However, one should note that the quantum effects that cause the creation and evaporation of the black hole singularity are expected to have scales of the order of the radial mass scale, thereby implying that the dominant features on any classically derived Penrose diagram are of the same scale as the quantum processes that modify those scales. The causal structure of the diagrams should serve as guidance for the construction of quantum processes in the space-time.

---